# CMOS Integrated Magnetless Circulators Based on Spatiotemporal Modulation Angular-Momentum Biasing

Ahmed Kord, *Student Member, IEEE*, Mykhailo Tymchenko, *Student Member, IEEE*, Dimitrios L. Sounas, *Senior Member, IEEE*, Harish Krishnaswamy, *Member, IEEE,* and Andrea Alù, *Fellow*, IEEE

*Abstract*— In this paper, we introduce the first integrated circuit (IC) implementation of spatiotemporally modulated angular-momentum (STM-AM) biased magnetless circulators. The design is based on a modified current-mode topology which is less sensitive to parasitics and relies on switched capacitors rather than varactors to achieve the desired modulation, thus reducing the circuit complexity and easing its chip-scale realization. We analyze the presented circuit and study its performance in the presence of inevitable non-idealities using an in-house so-called composite Floquet scattering matrix (CFSM) numerical method. We also validate the analysis with simulated and measured results using a standard 180 nm CMOS technology, showing good performance. Compared to previous discrete implementations of STM-AM circulators, the presented CMOS chip reduces the form factor by at least an order of magnitude and occupies a total area of only 36 mm$^2$.

## I. INTRODUCTION

MODERN communication networks are targeting tremendous increases in data rates to satisfy an ever-growing demand for faster and more efficient connectivity. Nevertheless, traditional sub 6 GHz bands of today's systems have become crowded with plenty of indoor and outdoor services, thus making it difficult to satisfy the new throughput requirements. To overcome this problem, the consensus over the past few years has been to shift to higher frequencies, where wideband unlicensed spectrum is still available. However, electromagnetic (EM) waves at such frequencies are less capable of penetrating obstructions, which reduces the area covered by each base station, complicates the tracking of fast-moving objects, and incurs other challenges [1], [2]. While many of these issues have been addressed over the past decade thanks to numerous developments by both the academic and industrial communities, the cost of deploying and maintaining the operation of such high-frequency systems remains very high compared to that of the current wireless infrastructure. Furthermore, all these systems are exclusively half-duplex, employing either frequency or time division diplexing to achieve bi-directional communication, therefore limiting the maximum transmission rate to only half of the network capacity. For this reason, full-duplex communication has been being investigated recently. In such a system, both the transmitter (TX) and the receiver (RX) operate simultaneously on the same frequency, which, in principle, doubles the spectral efficiency, or equivalently, reduces the operational cost by halving the required resources. Furthermore, full-duplex radios would provide a solution to many problems at the network layer such as hidden terminals, high end-to-end latency, fairness and congestion [3]-[9]. One major challenge in such systems, however, is that they require non-reciprocal components, namely circulators, at the radio-frequency (RF) front-end interface to mitigate the strong self-interference between the TX and RX nodes of each transceiver. Thus far, these components are almost exclusively based on magnetic biasing of rare-earth ferrite materials which must be of high crystal purity to achieve good performance. This, in turn, results in bulky and expensive devices, significantly lowers the yield of large scale production, and prohibits the ubiquitous use of magnetic circulators in the vast majority of commercial systems. To overcome these problems, magnetless implementations of such components have been pursued over several decades. However, despite significant research efforts, none of the previous works managed to achieve this goal while, at the same time, maintaining high performance that could satisfy all the necessary requirements of practical full-duplex radios [9]-[22]. For instance, active circulators which rely on the intrinsic non-reciprocal properties of transistors, have been broadly investigated [9]-[14], but they suffer from a

Manuscript received September 28, 2018. This work was supported in part by the Qualcomm Innovation Fellowship, in part by the IEEE Microwave Theory and Techniques Society Graduate Fellowship, in part by the Air Force Office of Scientific Research, in part by the Defense Advanced Research Projects Agency, in part by Silicon Audio, in part by the Simons Foundation, and in part by the National Science Foundation. (*Corresponding author: Andrea Alù*.)
A. Kord and M. Tymchenko are with the Department of Electrical and Computer Engineering, University of Texas at Austin, Austin, TX 78712, USA.
D. L. Sounas is with the Department of Electrical and Computer Engineering, Wayne State University, Detroit, MI 48202, USA.
H. Krishnaswamy is with the Department of Electrical Engineering, Columbia University, New York, NY 10027, USA.
A. Alù is with the Department of Electrical and Computer Engineering, University of Texas at Austin, Austin, TX 78712 USA, and also with the Advanced Science Research Center, City University of New York, New York, NY 10031, USA. (e-mail: aalu@gc.cuny.edu)

fundamentally large noise figure, limited power handling, and poor linearity [16]. Passive non-linear approaches have also been explored but they lead to inherent signal distortions and their operation is limited to a specific range of input intensities [17]. Therefore, neither of these approaches have been successfully commercialized. On the other hand, several attempts to build full-duplex radios without a circulator, by relying on asymmetric reciprocal interfaces such as electrical balance duplexers (EBDs) [23], [24] or by using multiple antennas, have also been pursued. Nevertheless, EBDs are theoretically limited to 3 dB insertion loss (>4 dB in practice) which weakens the argument of full-duplexing and makes it not worthy. Also, antenna cancellation, not only results in sensitivity issues to the antennas' placement, but it also makes full-duplexing less attractive compared to other communication paradigms such MIMO radios that can similarly double the throughput using multiple antennas but with less complexity.

Recently, it was shown that linear periodically time-varying (LPTV) circuits can overcome the limitations of all previous approaches and achieve high-performance magnetless non-reciprocity at low cost and small size, simultaneously [25]-[56]. In particular, [31] introduced the first discrete implementation of magnetless circulators based on STM-AM biasing of three coupled resonators, thus yielding large isolation in a compact footprint compared to magnetic devices. This work has raised significant interest in the microwave and circuit communities drawing attention to the exciting opportunities offered by LPTV networks in realizing non-reciprocal components without magnets, hence it was followed by numerous designs using similar concepts. For instance, [33] relied on staggered commutation of $N$-path filters to realize a highly miniaturized gyrator, which when embedded in a loop of reciprocal phase shifters yields the operation of a circulator. This was the first CMOS demonstration of this concept and it was followed by other contributions enhancing the performance of many metrics [33]-[38]. Specifically, [36] proposed a gyrator based on modulating the conductivity of a transmission line (TL), thus leading to a circulator with much wider isolation bandwidth and allowing to reduce the modulation frequency to one-third of the fundamental harmonic. A state-of-the-art implementation of this concept was presented in [38]. On the other hand, [40] presented an ultra-wideband circulator operating from 200 KHz to 200 MHz using sequentially switched co-axial cables. A miniaturized implementation of the same concept was presented in [41] using a 0.2 μm GaN HEMT technology. One challenge with [33]-[41], however, is the inherent *asymmetry* of the structures therein, which increases their sensitivity to inevitable random variations in practical systems. For instance, clock jitter and synchronization errors produced by an actual phase-locked loop (PLL) circuitry can degrade the overall isolation and the RX noise figure compared to their values under ideal conditions. A digitally modulated RF signal with a finite bandwidth and a high peak-to-average power can cause similar problems, especially when both the TX and RX signals are simultaneously fed to the circulator, as would be the case in real-life applications. Also, impedance variation at any of the circulator's ports, especially at the antenna (ANT) terminal, can be challenging to tackle in real time and may require complicated mixed-signal techniques that would impose a restriction on the maximum power handling. In order to overcome these critical problems, a cyclic-symmetric implementation that mimics the operation of magnetic-biased devices and maintains the rotational symmetry of ferrite cavities is highly desirable. Towards this goal, [45] presented a cyclic-symmetric STM-AM magnetless circulator resulting in Watt-level power handling for the first time. Also, [46] presented a pseudo-linear time-invariant magnetless circulator with remarkable performance nearly in all metrics, including a 0.8 dB insertion loss (after de-embedding the balun losses) and 2.5 dB noise figure, the lowest among all magnetless circulators presented to-date. Furthermore, [50] proposed a technique to increase the 20 dB isolation bandwidth to 14% of the center frequency. Despite the remarkable performance in [45]-[50], the designs therein were all based on printed-circuit board (PCB) technology and off-the-shelf discrete components, which limits their size and cost reduction and prohibits their large-scale production. In order to overcome this problem, we present in this paper the first CMOS implementation of STM-AM magnetless circulators.

This paper is organized as follows. In Sec. II, we discuss the physical principles of STM-AM biasing and explain the challenges of integrating the conventional voltage- and current-mode topologies using standard CMOS technologies. In Sec. III, we present a modified current-mode architecture based on switched capacitors with binary frequency shift keying (BFSK) clocking scheme that eases chip-scale implementation. In the same section, we investigate the impact of inevitable parasitics and clock non-idealities on the performance using an in-house numerical method, showing the robustness of the proposed circuit against such errors. In Sec. IV, we validate our analysis with simulated and measured results using a standard 180 nm CMOS process, showing good performance. Finally, we draw our conclusions and provide an outlook on future directions in Sec. V.

## II. SYNTHETIC STM-AM BIASING

Single-ended (SE) implementations of STM-AM magnetless circulators were initially presented in [31], [45] based on connecting three bandpass or bandstop resonators either in series or in parallel and modulating their oscillation frequencies with a particular phase pattern through varactors. The circuit in [45], in particular, was the first magnetless circulator to achieve Watt-level power handling and superior linearity that exceeds by orders of magnitude all previous active approaches. Nevertheless, [45] suffered from strong intermodulation products (IMPs) at all ports, due to mixing between the RF and the relatively low-frequency modulation signals, which limited the lowest possible insertion loss and, consequently, noise figure to about 3 dB at best. In order to overcome this problem, [46] proposed differential implementations of STM-AM circulators based on connecting two identical SE circuits either in series through baluns or in parallel by tying their respective terminals, resulting in what we called voltage- and current-mode architectures, respectively, in analogy with passive mixers. In order to explain the challenges associated with integrating either of these circuits and to develop the new circuit presented in Sec. III, we begin by inspecting the conventional current-mode topology depicted in Fig. 1. This circuit consists of two single-ended circulators, each of which contains three series $LC$ tanks connected in a wye topology and modulated through variable capacitors as follows

$$f_n = \begin{cases} f_0 + u(t-(n-1)T_m/3), & n=1,2,3 \quad \forall SE1 \\ f_0 - u(t-(n-1)T_m/3), & n=4,5,6 \quad \forall SE2 \end{cases} \quad (1)$$

where $n$ is the tank number, $f_0$ is the static oscillation frequency of all tanks, $u(t)$ is a periodic function that determines the modulation scheme, $T_m = 1/f_m$ is the modulation period, and $SE1$ and $SE2$ are the constituent SE circulators. In general, the port currents in this circuit can be written as a superposition of its eigenstates, i.e. the common mode $I_c$, the clockwise mode $I_+$, and the counter clockwise mode $I_-$ (see Appendix A). Assuming that the common mode is zero around the center frequency, the port currents $I_k$ can be written as follows

$$I_k = I_+ e^{+j(k-1)2\pi/3} + I_- e^{-j(k-1)2\pi/3}, \quad (2)$$

where $k$ is the port number. Due to symmetry, an incident wave at any port, say port 1, excites $I_+$ and $I_-$ with the same amplitude and phase, hence the output signals at the other ports 2 and 3 become identical. Equation (1) provides a preferred sense of precession in the counter-clockwise direction, based on the port definitions of Fig. 1, since the phases of the modulation signals increase in that direction. Notice that, by definition, the phases of $I_+$ and $I_-$ increase in opposite directions, therefore (1) lifts their degeneracy and, consequently, makes them oscillate at different frequencies, i.e., $f_+$ and $f_-$, respectively. Also, the amount of splitting $\Delta f = (f_+ + f_-)/2$ depends on the characteristics of $u(t)$. For instance, [46] assumed $u(t) = u_0 \cos(2\pi f_m t)$, which resulted in $f_\pm \approx f_0 \mp f_m$ and led to the normalized magnitude and phase spectrums of $I_\pm$ shown in Fig. 2(a) and Fig. 2(b), respectively. At the center frequency $f_0$, the amplitudes of $I_\pm$ are equal while the phases are exactly opposite, i.e., $I_\pm = I_0 e^{\pm j\alpha}$, therefore, (2) simplifies to

$$I_k\big|_{f=f_0} = 2I_0 \cos\left[(k-1)\frac{2\pi}{3} + \alpha\right]. \quad (3)$$

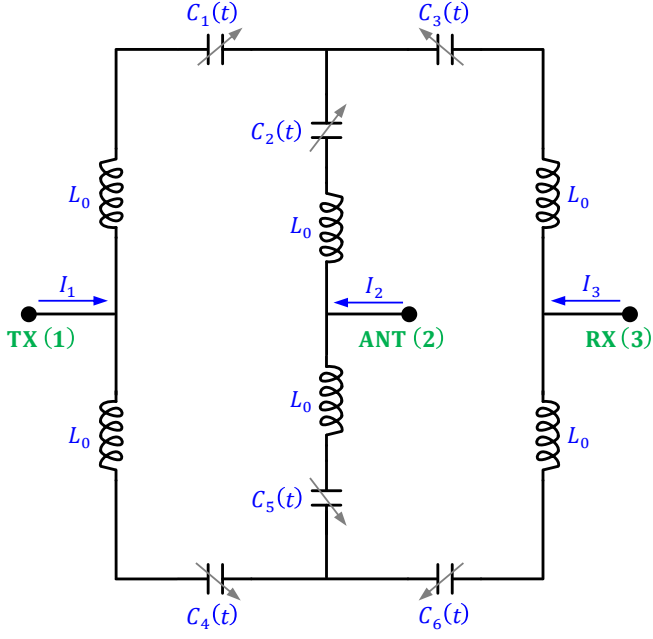

Fig. 1. Differential current-mode STM-AM circulator consisting of two SE wye topologies connected in parallel.

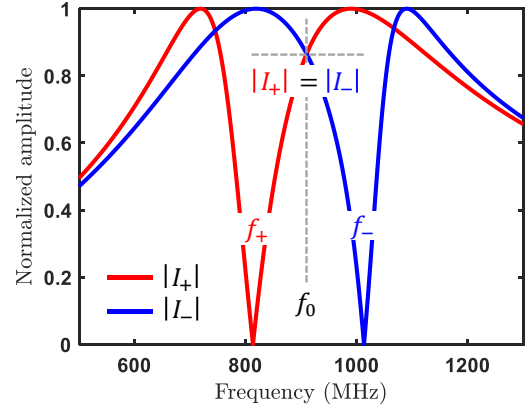

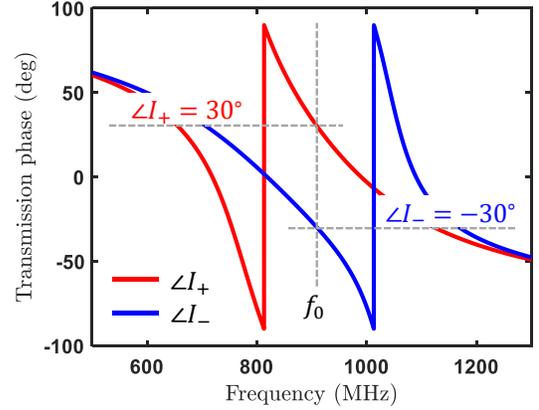

Fig. 2. Spectrums of the rotating modes $I_+$ and $I_-$. (a) Normalized amplitude. (b) Phase.

By controlling the amplitude and frequency of $u(t)$, $\alpha$ can be designed to be 30 deg, which when substituted in (3) results in $I_3 = 0$ and $I_1 = -I_2 = \sqrt{3}I_0$, thus isolating port 3 from excitations at port 1 and transmitting all the input power to port 2. As the input frequency deviates from $f_0$, however, the amplitudes and the phases of $I_\pm$ become different, thus reducing the isolation (IX) and increasing both the insertion loss (IL) and the return loss (RL), which results in the typical dispersive S-parameters of circulators.

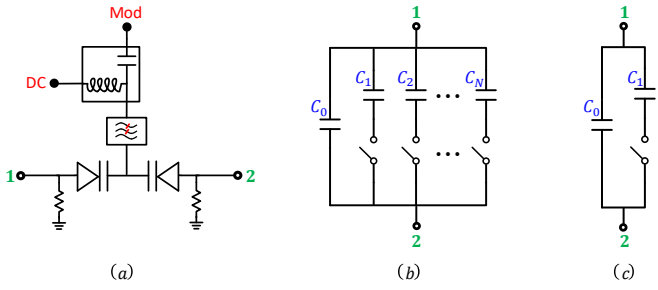

Fig. 3. Implementation of the variable capacitor $C(t)$ using: (a) Varactors. (b) $N$ periodically switched capacitors. (c) BFSK switched capacitor.

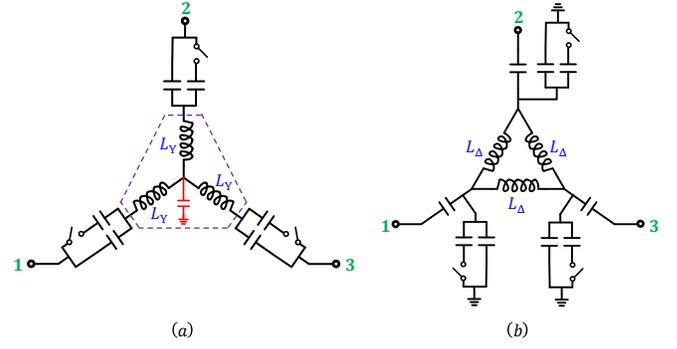

Fig. 4. (a) Conventional wye topology depicting a shunt parasitic capacitance at the central node. (b) Modified wye topology where the inductors are transformed into a delta connection and the switched capacitors are connected in shunt rather than in series.

The sinusoidal modulation scheme described above, i.e., $u(t) = u_0 \cos(2\pi f_m t)$, was achieved in [46] by replacing each variable capacitor $C(t)$ in Fig. 1 with a pair of common-cathode varactors and injecting the required DC bias and the modulation signals through low-pass filters to the common-cathode node, as shown in Fig. 3(a). While such implementation did indeed result in excellent performance using discrete components as demonstrated in [45], [46], it actually suffers from several issues that complicate its CMOS integration. First and foremost, the varactors' low-pass filters require using multiple lumped elements with considerably large values, e.g., inductors are in the order of hundreds of nH, which cannot be integrated on a chip. Furthermore, the required modulation index, i.e., capacitance variation, to achieve sufficient isolation in STM-AM circulators can in general be quite large, i.e., in the order of 50%. While discrete varactors can achieve such value and maintain a high quality factor at the same time, it is impossible to attain similar performance using standard CMOS components. These problems can be overcome by replacing the varactors with $N$ periodically switched capacitors as shown in Fig. 3(b). Such implementation not only eliminates the complicated biasing networks of varactors and permits arbitrarily large modulation indices, but it also replaces the sinusoidal signals with digital clocks which are easier to generate in multiple phases and are less sensitive to amplitude noise. The downside, however, is that switched capacitors can only synthesize *quantized* modulation schemes regardless of how large $N$ might be, which leads to finite IMPs and imposes a restriction on the lowest possible IL, even with ideal lossless components. Fortunately, if these products are sufficiently far from the center frequency, i.e., $f_m \ll \text{BW}/2$, then they can be rejected by the junction's bandpass resonance itself. Under this condition, even an abrupt periodic variation of $C(t)$ between two values, say $C_0$ and $C_0 + C_1$ as depicted in Fig. 3(c), can lead to excellent performance. Obviously, this requires the unmodulated resonance of the circulator to be of bandpass type, which excludes the voltage-mode topology and justifies the focus on the current-mode architecture thus far.

In IC designs, the parasitics can also be of major concern. For instance, shunt parasitics at the central node of the wye topology may reduce the coupling between the constituent *LC* tanks and isolate them from one another, thus degrading the overall performance. This problem can be overcome by transforming the wye inductors highlighted in Fig. 4(a) into three new delta inductors, as depicted in Fig. 4(b). Furthermore, the switched capacitors in Fig. 4(b) are connected in shunt rather than in series in order to reduce the swing across them when they are ON, which improves the circulator's power handling and linearity. Notice that these metrics are still limited by the OFF state non-linearities, but these are typically much weaker in most IC technologies. It is also worth mentioning that any shunt parasitic capacitance at the terminals of the inductance loop of Fig. 4(b) can be absorbed in the switched capacitors, thus easing the integration of this circuit.

## III. CMOS IMPLEMENTATION

Fig. 5(a) shows the complete schematic of the proposed CMOS STM-AM circulator based on the modified wye topology depicted in Fig. 4(b). Specifically, the circuit consists of two SE circulators connected in parallel, each of which contains three inductors $L_0$ connected in a loop, the terminals of which are coupled to the RF ports through a capacitor $C_0$ and to ground through a parallel combination of a static capacitor $C_g$ and a switched capacitor $C_m$. Fig. 5(b) also shows the timing diagram of the switching clocks $s_n$, which all have the same frequency $f_m = 1/T_m$ and they are related together as follows

$$s_n(t) = s_{n-1}(t - nT_m/3), \quad n = 1,2,3 \quad (4)$$

$$s_n(t)\big|_{n=4,5,6} = s_n(t - T_m/2)\big|_{n=1,2,3}. \quad (5)$$

where $s_0(t)$ is a square wave function. In the special case of a 50% duty cycle, as we assume in this paper, (5) simplifies to $s_{4,5,6} = \bar{s}_{1,2,3}$, i.e., the clocks of the two SE circuits are inverted versions of each other. The modulation scheme described by (4) and (5) is equivalent to substituting

$$u(t) = \begin{cases} 0, & 0 < t < T_m/2 \\ u_0, & T_m/2 < t < T_m \end{cases} \quad (6)$$

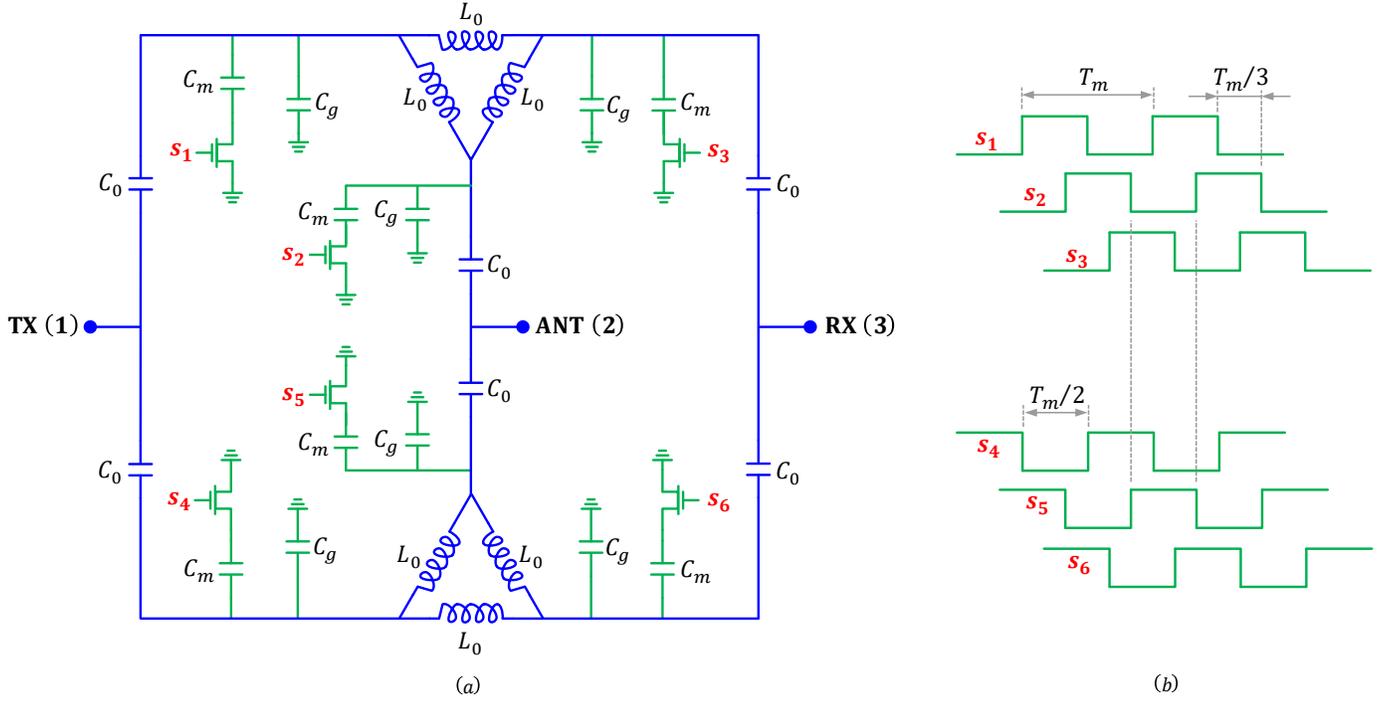

Fig. 5. CMOS STM-AM circulator based on a modified current-mode topology. (a) Complete schematic. (b) Timing diagram of the switching clocks.

in (1), where $u_0$ is a constant that depends on the value of $C_m$. As a result, the oscillation frequencies $f_n$ change periodically between two values, i.e., $f_0$ and $f_0 - u_0$. Interestingly, this is strikingly similar to the well-known BFSK modulation scheme commonly used in digital communication systems, hence we use the same terminology and refer to the timing diagram of Fig. 5(b) as BFSK clocking. This analogy suggests that more advanced modulation schemes may improve certain metrics of the circulator, e.g., reducing spurious emission or increasing the isolation bandwidth, yet this is beyond the scope of this paper and will be the subject of future publications.

Table I summarizes the values of all elements and parameters used in obtaining the numerical results to be discussed in this section. The values used in obtaining the simulated and measured results, which will be discussed in Sec. IV, are also listed. The inductors $L_0$ and the capacitors $C_0$ were chosen such that the unmodulated circuits of Fig. 5(a) would resonate at the target frequency of 915 MHz. Then, the modulation frequency $f_m$ was set to 106 MHz, i.e. 11.6% of the RF center frequency. Such a small value permits feeding the modulation signals externally from regular testbench generators as was the goal in this paper. In future designs, however, the modulation sources should be integrated with the circulator on the same chip to allow using higher $f_m$ if necessary. Recall that higher $f_m$ would push the IMPs further away from the center frequency, thus improving the spurious emission of the circulator and reducing its IL, as explained in Sec. II. It is also worth mentioning that the specific value of 106 MHz was chosen to avoid any high-order harmonics of the modulation signals landing inside the desired BW centered around 915 MHz. Based on this choice of $f_m$, the capacitors $C_g$ and $C_m$ were chosen to maximize the achieved IX at 915 MHz. Finally, the transistors were sized at W/L = (150×2µ)/340n to minimize their on-resistance.

### A. Impact of Resistive Losses

In contrast to balanced duplexers, the IL of the proposed circuit is not theoretically limited to 3 dB and it only depends on the resistive parasitics, mainly $r_{on}C_{off}$ of the switches and $Q_0$ of the inductors. Fig. 6 shows the impact of changing these parameters on the numerical S-parameters obtained using an in-house so-called CFSM numerical method (see Appendix B). Intuitively, larger $Q_0$ and smaller $r_{on}C_{off}$ leads to better RL and IL, as indeed depicted in Fig. 6. For instance, $r_{on}C_{off} = 0.4$ psec and $Q_0 = 80$ results in RL and IL of about 12.8 dB and 1.8 dB, respectively. While such value of $r_{on}C_{off}$ can be achieved using deep-submicron CMOS technologies, $Q_0 = 80$ is only possible using off-chip inductors. The quality factor of on-chip inductors, on the other hand, is in the order of 10~20 at the first few GHz, which makes achieving low IL more challenging and limits the miniaturization of the proposed circuit. For example, Fig. 6(a) and Fig. 6(b) show that $r_{on}C_{off} = 0.4$ psec and $Q_0 = 20$ results in RL and IL of 8.6 dB and 5.2 dB, respectively. This is because STM-AM circulators, similar to their magnetic counterparts, are resonant circuits, therefore their IL is sensitive to the resonance unloaded quality factor. Therefore, we will assume that the inductors in this paper are off-chip. Another reason to motivate this assumption is that the value of $L_0$ is quite large, i.e., 30 nH, which is difficult to be implemented on a chip. On the other hand, Fig. 6(c) shows a rather non-trivial dependence of the maximum possible IX on $r_{on}C_{off}$ and $Q_0$. In fact, for each $r_{on}C_{off}$ there is a specific $Q_0$ at which a near-perfect IX can be achieved. For example, the optimal $Q_0$ for $r_{on}C_{off}=0.4$ psec is about 64. Either a smaller or even a larger $Q_0$ would change the amount of splitting of the rotating modes (see Fig. 2) and as such the maximum achieved IX is reduced. In practice, however, it is sufficient to maintain the in-band IX larger than 20~30 dB, i.e., the level at which ANT reflections due to dynamic impedance mismatches begin to dominate the leakage at the RX port. Fig. 6(c) shows that such level is maintained for $Q_0$ ranging from 20 to more than 100, thus showing that the IX of the proposed circuit is insensitive to the inevitable random variations of the losses.

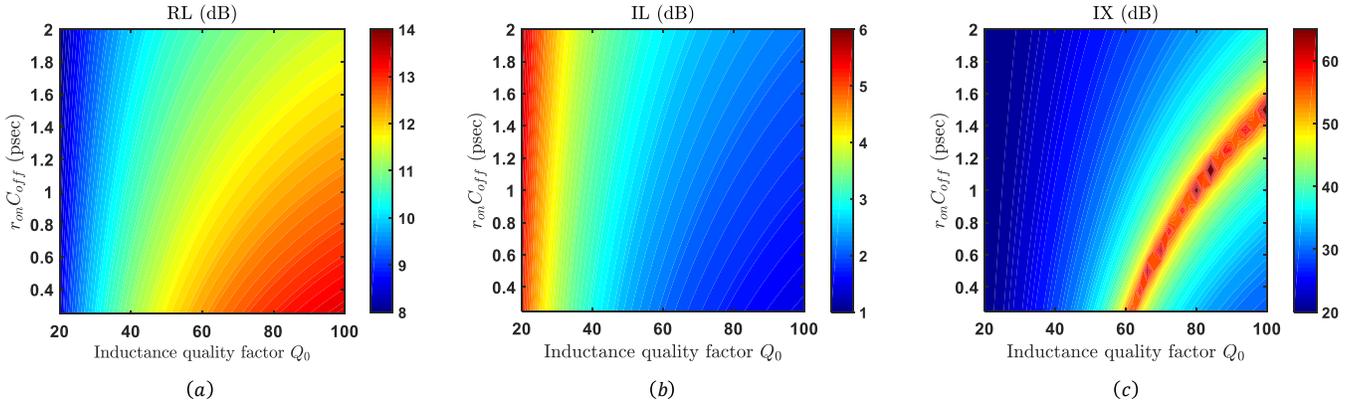

Fig. 6. Impact of switch parasitics $r_{on}C_{off}$ and inductance quality factor $Q_0$ on the S-parameters. (a) RL. (b) IL. (c) IX.

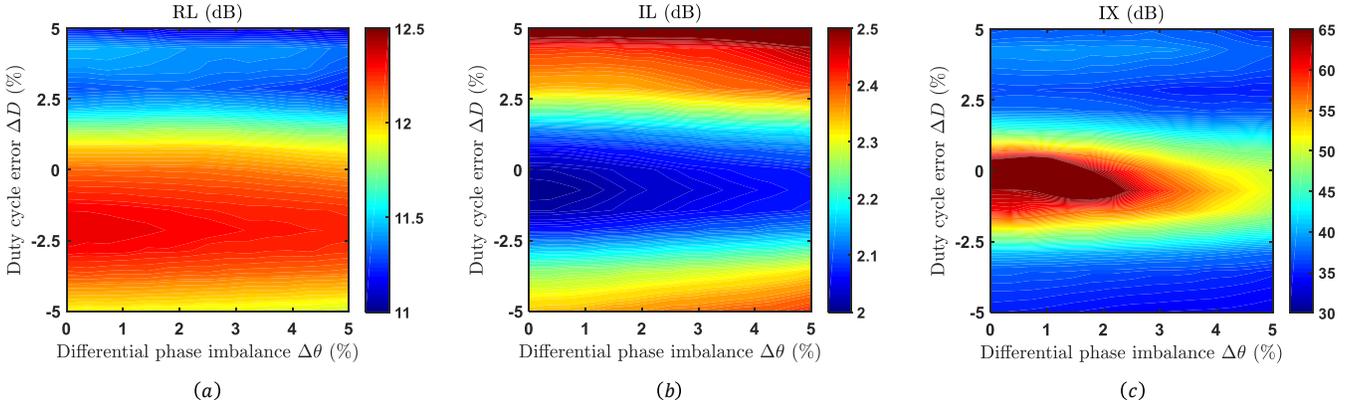

Fig. 7. Impact of duty cycle error $\Delta D$ and differential phase imbalance $\Delta\theta$ on the S-parameters. (a) RL. (b) IL. (c) IX.

*B. Impact of Clock Non-idealities*

The clock signals in this paper are assumed to be generated externally using three phase-locked RF sources, thus allowing precise control of the duty cycle and the relative phases. In real applications, however, these clocks will be generated on-chip, which can be easily achieved using digital circuitry, but they will also be prone to phase noise, jitter, and duty cycle errors. In order to quantify the sensitivity of the proposed circulator to such errors, Fig. 7 shows the numerical S-parameters versus fractional errors $\Delta D$ and $\Delta\theta$ in the duty cycle and the phase difference between the constituent SE circulators, respectively. Clearly, the network is quite tolerant to differential phase imbalance, allowing $\Delta\theta \approx 2$ % (for $\Delta D=0$) without any noticeable degradation in its performance. The requirements on the duty cycle errors, however, turn out to be more stringent. For instance, $\Delta D \approx +0.7$ % (expansion of the switching on-time leading to overlap periods between the clocks of the single-ended circuits) with $\Delta\theta=0$ reduces the IX by about 20. In contrast, $\Delta D \approx -0.7$ % (compression of the switching on-time) leads to only 12 dB IX degradation and even a modest improvement of RL and IL compared to the ideal case, clearly indicating that it is important to ensure that the switches in each of the circulators do not overlap.

IV. RESULTS

Based on the numerical findings of Sec. III, an IC STM-AM circulator was designed for operation at 915 MHz using a standard 180 nm CMOS technology. Table I summarizes the values of all elements and parameters used in obtaining the simulated and measured results to be discussed in this section. Notice that these values are different than their numerical counterparts due to the additional parasitics of the physical elements. In particular, $C_g$ was entirely eliminated since the shunt parasitic caps of the off-chip inductors were sufficient to realize it without requiring any additional on-chip capacitors. Fig. 8(a) also shows a picture of the chip layout where the active silicon area occupied by the circulator is only 0.2 mm2.

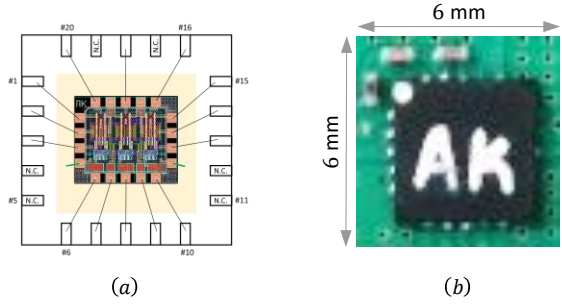

Fig. 8. (a) Picture of the chip layout. (b) Top-view photograph of the packaged chip mounted on a PCB. Three more inductors are placed at the bottom side of the PCB.

TABLE I
VALUES OF ALL DESIGN PARAMETERS USED IN OBTAINING THE NUMERICAL, SIMULATED, AND MEASURED RESULTS PRESENTED IN THIS PAPER.

| Element\Value | Numerical | Simulated | Measured |
|---|---|---|---|
| $f_m$ (MHz) | 106 | 106 | 106 |
| $L_0$ (nH) | 40 | 30 (Coilcraft) | 30 (Coilcraft) |
| $C_0$ (fF) | 900 | 450 (MIM) | 450 (MIM) |
| $C_g$ (fF) | 900 | N/R | N/R |
| $C_m$ (fF) | 1200 | 2×800 (MIM) | 2×800 (MIM) |
| $V_{dd}$ (Volt) | N/R | 3.3 | 3.3 |
| $W/L$ (μm/μm) | N/R | (150×2)/0.34 | (150×2)/0.34 |

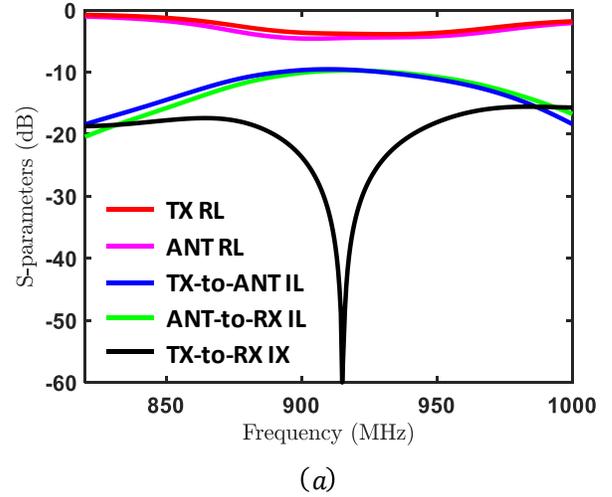

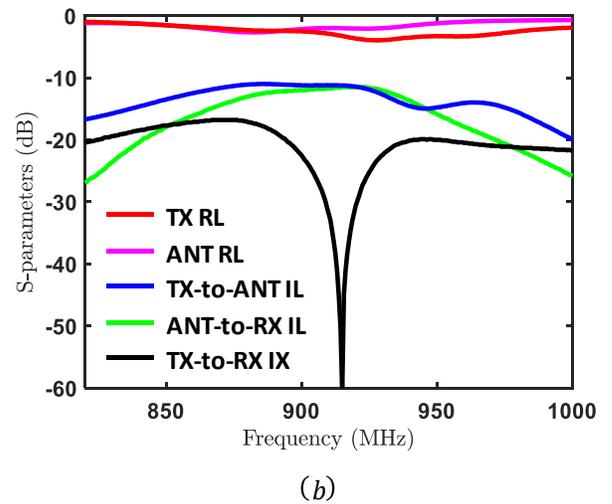

Fig. 9. *S*-parameters before matching. (a) Simulated. (b) Measured.

The chip was also mounted in a 5×5 mm2 QFN package and connected to off-chip inductors on a PCB for testing, as shown in Fig. 8(b). The total form factor is 6×6 mm2 which is at least one order of magnitude smaller than the discrete implementations of the conventional voltage- and current-mode architectures presented in [46]-[50].

Fig. 9(a) and Fig. 9(b) show a comparison between the simulated and measured *S*-parameters of the circulator, respectively, when its ports are all terminated with 50 Ohm impedances. Obviously, large TX-to-RX IX of more than 60 dB is achieved at a center frequency of 915 MHz in both cases, however, the circulator also exhibits significant transmission losses. Specifically, the measured TX-to-ANT IL and ANT-to-RX IL are 11 dB and 11.5 dB, respectively. These large values are not actually due to resistive or even spurious emission losses but they are rather an artifact resulting from strong impedance mismatch as evident by the significant TX RL and ANT RL of 3.8 dB and 2.3 dB, respectively. This matching problem was not predicted in our initial simulations since the custom-designed electrostatic discharge (ESD) protection circuitry was not taken into account. After taking the measurements, however, we realized that the ESD diodes introduce significant shunt parasitics at the ports, which lead to strong reflections. When the ESD circuit was accurately modeled and incorporated in post-layout simulations, the results agreed well with the measurements and the matching problem was similarly observed. Specifically, the simulated TX-to-ANT IL and ANT-to-RX IL are both found to be 9.6 dB, while the TX RL and ANT RL are 4.3 and 3.8 dB, respectively. This problem was fixed in simulations by adding proper matching networks at the circulator's ports to tune out the ESD parasitics, thus yielding the results depicted in Fig. 10(a). As we can see, the TX RL and ANT RL both improve to 12.5 dB and, subsequently, the TX-to-ANT IL and ANT-to-RX IL reduce to about 3 dB. Also, the center frequency is shifted down to about 912 MHz. These results were experimentally validated by using three external reconfigurable impedance tuners, yielding the measured results shown in Fig. 10(b). The measured TX RL and ANT RL are 11 dB and 13 dB, respectively, the measured TX-to-ANT IL and ANT-to-RX IL are both 4.9 dB, and the center frequency is about 910 MHz. Furthermore, the simulated and measured 20 dB IX BW are 3.4% (31 MHz) and 2.4% (22 MHz), respectively. It is worth mentioning that the IL can be reduced to less than 3 dB, as predicted by the numerical analysis of Sec. III, by further optimizing the chip and the PCB layouts and by better designing the IC supplementary blocks, namely, the modulation path buffers and the ESD protection circuit, which is the focus of ongoing investigations and shall be reported in future publications. Fig. 11 also shows a comparison between the simulated and measured transmission phases, which are all quite linear within the BW of interest (highlighted in grey color). Surprisingly, this metric is overlooked in several recent works although it might be critical in real systems to avoid anomalous dispersion of digitally modulated RF signals.

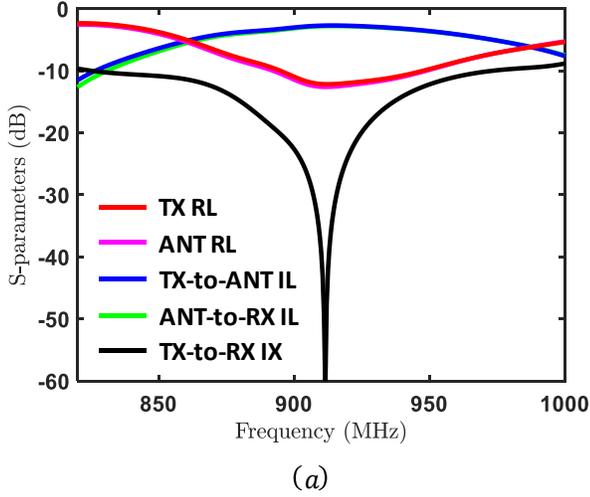
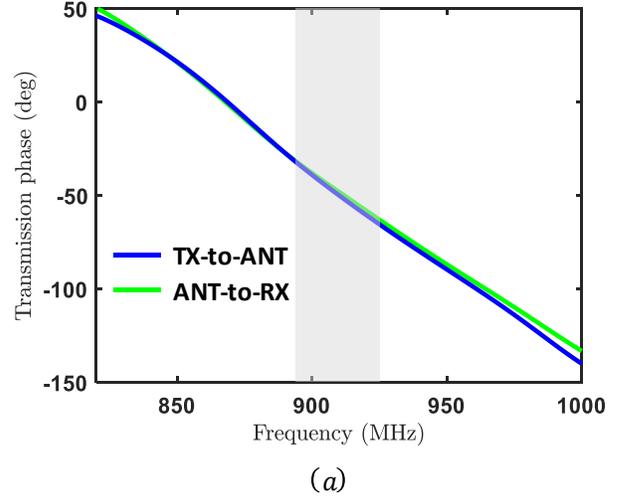
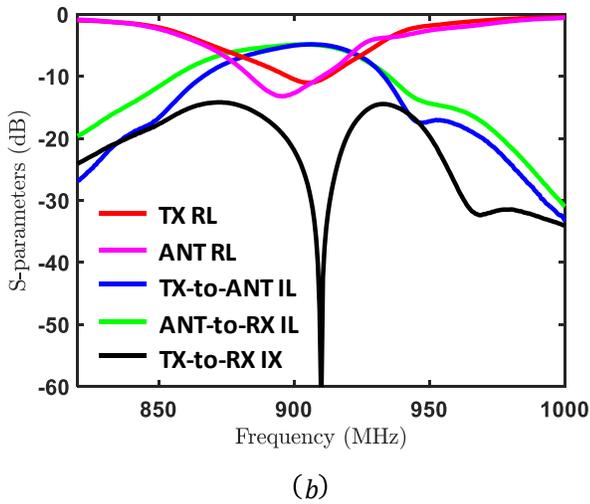
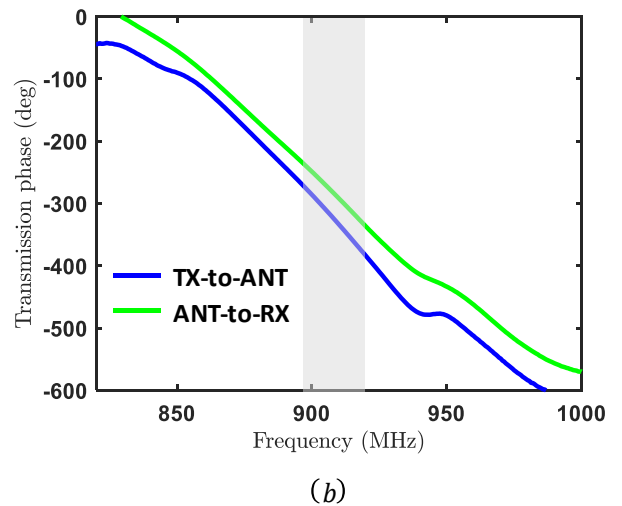

Fig. 10.  *S*-parameters after matching. (a) Simulated. (b) Measured.

Fig. 11.  Transmission phase. (a) Simulated. (b) Measured.

Fig. 12(a) and Fig. 12(b) also show a comparison between the simulated and measured spectrums, respectively, at both the ANT and the RX ports for a monochromatic TX excitation at 910 MHz. Results are normalized to the input power for simplicity. The fundamental harmonic at the ANT port exhibits a simulated IL of about 3 dB and a measured IL of 4.8 dB, in agreement with the *S*-parameters. However, the fundamental harmonic at the RX port depicts a simulated and a measured IX of 37 dB and 42 dB, respectively, instead of 65 dB, as predicted by the *S*-parameters. This is simply because the input frequency of the TX signal is not aligned with the IX dips in Fig. 10. In addition to the fundamental component, the spectrums at both the ANT and RX ports are also contaminated with many spurs. These are essentially higher-order harmonics of the modulation signals and IMPs resulting from their mixing with the TX input signal. The largest measured IMPs are those at 809 MHz and 1021 MHz, which are –19 dBc and –22 dBc, respectively, while the simulated value of both is –22 dBc. In general, these products depend strongly on the balance of the differential circuit, hence they are more sensitive to layout asymmetries than other metrics such as the *S*-parameters. Also, as mentioned earlier, the modulation signals in this paper are generated externally from regular testbench sources which are configured to generate three 120 deg phase shifted sinusoidal signals at frequency $f_m$. These signals are then fed to the chip through ESD protected pins and buffered through a series of inverters that transform them into digital clocks which are eventually applied to the gates of the switches. As one may expect, these clocks are not perfect square waves and they may incur considerable amplitude variations which enrich their frequency content. This, in turn, explains why the beat frequency in Fig. 12(a) and Fig. 12(b) is smaller than $f_m$. As one may expect, such spurious emission must be further suppressed in order to comply by the spectral mask regulations of commercial systems and to avoid interference with adjacent channels. This can be achieved by using a combination of various solutions as follows. First, the modulation signals ought to be generated on chip rather than feeding them from external sources, which would ease controlling their frequency content and ensure that they are exactly square waves with a fundamental frequency $f_m$. This also permits increasing $f_m$ in order to push the IMPs further away from the BW of interest such so that they exhibit more attenuation by the natural bandpass response of the resonant wye junctions themselves. Another straightforward solution is to add bandpass filters at the ports of the circulator to reject the out-of-band spurs. However, this is not preferred in practice as it increases the overall size and cost and degrades the overall IL. Other sophisticated solutions include using more advanced modulation schemes than the simple BFSK adopted in this paper or combining different circulators together with a particular phase shift between their modulation signals, yet these are beyond the scope of this paper and will be presented in future publications.

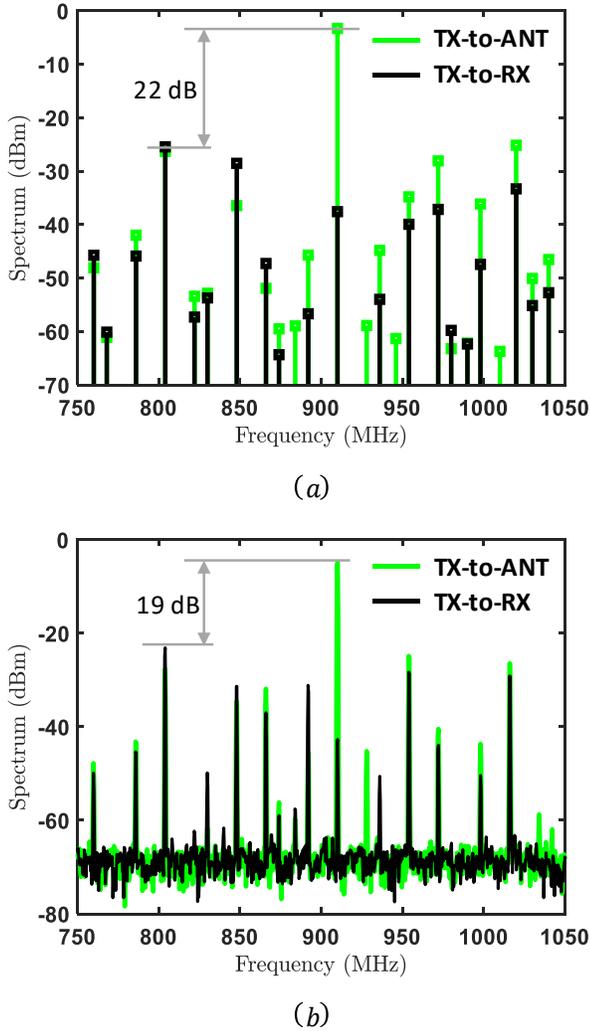

Fig. 12. Normalized spectrums at the ANT and RX ports for a monochromatic TX excitation at 910 MHz. (a) Simulated. (b) Measured.

Finally, the TX-to-ANT and ANT-to-RX third-order input intercept point (IIP3) are measured and found to be +6.1 dBm and +6 dBm, respectively. These values are, in fact, currently limited by the ESD protection circuitry which, as mentioned earlier, was not optimally designed. Therefore, should this issue be fixed in future designs, the IIP3 is expected to increase. Also, an RX noise figure (NF) of 5.2 dB was measured at the circulator's center frequency of 910 MHz. This is only 0.4 dB larger the ANT-to-RX IL, thus showing negligible degradation due to the modulation phase and amplitude noise.

Table II summarizes the presented measured results in comparison to previous works on magnetless circulators based on various technologies. Similar to [44]-[50], this work relies on cyclic-symmetric STM-AM biasing to achieve magnetless non-reciprocity, while [33] relies on $N$-path filtering, [37] and [40] rely on TL switching, and [14] and [15] rely on active techniques to realize a quasi-circulator. Out of these different works, [15] is the only one that achieves a transmission gain rather than an IL. However, this comes at the expense of a significantly large RX NF of 16 dB and a very low TX-to-ANT IIP3 of –11 dBm. As explained in the introduction, these are fundamental limitations of active circulators, therefore even though [14] does not report the values of these metrics, one can expect that it suffers from the same problems. Amongst all other works, the lowest IL and NF are both achieved by [46] based on the STM-AM architecture. However, this is a discrete implementation, therefore the size is quite large.

The presented work overcomes the size problem of [46] and, at the same time, the poor NF and IIP3 of [14] and [15] by introducing the first CMOS implementation of STM-AM circulators. Specifically, the total form factor herein is 36 mm2, i.e., $\lambda_0/50 \times \lambda_0/50$ where $\lambda_0$ is the free-space wavelength, which is about one order of magnitude smaller than [46]. Furthermore, only 35% of this area is occupied by actual elements while the remaining 65% is consumed by the package itself, thus allowing for further miniaturization in future designs. Notice that the much smaller size of [37] and [14] is mainly due to the fact that they are not packaged and partially because of the 25~30 times higher center frequency which reduces the wavelength and scales down the size of all elements accordingly. On the other hand, the presented work outperforms [14] and [15] in terms of NF and IIP3 because STM-AM circulators are essentially passive low-noise non-reciprocal circuits as explained earlier [51]. It is also worth mentioning that both TX-to-ANT IIP3 and ANT-to-RX IIP3 achieved in this work are comparable to the ANT-to-RX IIP3 of [33] and [37] but much lower than the TX-to-ANT IIP3 therein. This is mainly because the asymmetry of [33] and [37] gives them an advantage in terms of this metric, yet this is at the expense of increasing the sensitivity to random variations as explained in Sec. I. In order to enhance the linearity and power handling while maintaining the cyclic-symmetry, we are currently investigating a technique based on interconnecting multiple STM-AM circulators together with a particular phase shift between their modulation signals. However, this is beyond the scope of this paper and will be presented in future publications.

TABLE II
SUMMARY OF THE MEASURED RESULTS IN COMPARISON TO PREVIOUS WORKS.

| Metric\Reference | This work | [45] | [46] | [50] | [33] | [37] | [40] | [14] | [15] |
|---|---|---|---|---|---|---|---|---|---|
| Architecture | STM-AM | | | | $N$-path filtering | TL switching | | Active Circ. | |
| | Modified wye | SE | Diff. | Broad. | | | | | |
| Technology | CMOS 180 nm | Discrete | Discrete | Discrete | CMOS 65 nm | CMOS SOI 180 nm | Coaxial cables | CMOS 180 nm | CMOS 180 nm |
| Center freq. (GHz) | 0.91 | 1 | 1 | 1 | 0.75 | 25 | 0.1 | 30 | 2.4 |
| Modulation freq. (%) | 11.6 | 19 | 10 | 11 | 100 | 33.3 | 6 | N/R | N/R |
| 20 dB IX BW (%) | 2.4 | 2.4 | 2.3 | 14 | 4.3 | 18.4 | 200 | N/D | 9 |
| TX-to-RX IX (dB) † | 65 | 55 | 24 | 31 | 50 * | 18~20 | 40~70 | 16 | 76 |
| TX RL (dB) † | 11 * | 11.3 | 23 | 16.7 | N/A | 15 | 20 | 6.5 | 8 |
| ANT RL (dB) † | 13 * | 11.3 | 23 | 14.6 | N/A | 15 | 22 | 5.8 | 20 |
| TX-to-ANT IL (dB) † | 4.8 * | 3.3 | 0.8 ɤ | 4.2 | 1.7 | 3.3 | 9 | 4 | –23 # |
| ANT-to-RX IL (dB) † | 4.8 * | 3.3 | 0.8 ɤ | 4.2 | 1.7 | 3.2 | 8 | 7.5 | –12 # |
| Transmission phase (deg) | Linear | Linear | Linear | Linear | N/A | N/A | N/A | N/A | N/A |
| Max. IMP (dBc) | –19 | –11 | –29 | –22 | N/A | N/A | N/A | N/R | N/R |
| TX-to-ANT IIP3 (dBm) | +6.1 | +33 | +32 | N/A | +27.5 | N/A | N/A | N/A | –11 |
| ANT-to-RX IIP3 (dBm) | +6 | +33 | +32 | N/A | +8.7 | N/A | N/A | N/A | +0.6 |
| RX NF (dB) † | 5.2 | 4.5 | 2.5 | N/A | 4 | 3.3 | N/A | N/A | 16 |
| Power consum. (mWatt) | 64 ¥ | N/A | N/A | N/A | N/A | N/A | N/A | 15 | 145 |
| Size (mm2) | 36 | 143 | 286 | 480 | 25 | 2.16 ** | N/R | 0.365 ** | 3.2 ** |

N/A: Not available.    † At center frequency.    * With external impedance tuners.    ** Bare die without packaging.
N/R: Not relevant.    ɤ Baluns are de-embedded.    # Transmission gain.
N/D: Not defined.    ¥ Not including power consumed in generating the clocks.

The largest 20 dB IX BW is achieved by [40]. However, this is based on coaxial cables which occupy a huge area, obviously. Should this design be built using discrete or integrated components for a fair comparison with other works, the BW is expected to reduce and the modulation frequency will also increase dramatically. This is because the underlying TL switching architecture of this work requires electrically large delay lines to achieve a wide BW. Notice that the BW of [14] is undefined since the maximum TX-to-RX IX is already smaller than 20 dB. Also, IX in [37] is less than 20 dB, however, it is maintained between 18 dB and 20 dB over a wide frequency range, hence this range is approximately defined as the 20 dB IX BW. The presented work, on the other hand, is narrowband, similar to [45] and [46]. Nevertheless, the 2.4% BW can be increased more than threefold by cascading the chip with bandpass filters tailored to exhibit a certain output impedance, as explained in [50].

The presented work also consumes a relatively low power, only 64 mWatt, thanks to the low modulation frequency of the STM-AM architecture. This value, however, does not include the power consumed in generating the modulation signals themselves. But since the presented design relies on switched caps rather than varactors, the modulation signals are simply digital clocks which would be necessarily available in a fully integrated full-duplex system to serve other mixed-signal and RF blocks. Although frequency multipliers and phase shifters may still be required to provide the required $f_m$ and 120 deg phase shifts in particular, these are all digital operations that can be performed with negligible power consumption. Therefore, the reported 64 mWatt can be fairly compared to that of active circulators, i.e., [14] and [15]. Interestingly, [15] consumes more than twice this power. This is mainly because the design therein was intended to provide a transmission gain rather than an IL. On the other hand, [14] consumes very small power but the performance of all other metrics are worse than the presented CMOS STM-AM circulator.

It is also worth highlighting that the linear dispersion of the in-band transmission phase has been confirmed only for STM-AM circulators including the presented CMOS implementation, but not for other works. If such phase is non-linear, however, the circulator would result in anomalous dispersion of digitally modulated RF signals, which would increase the bit error rate in real full-duplex systems. Similarly, amongst all pertinent implementations, spurious emission has been investigated only in those based on the STM-AM architecture. Ensuring weak emission is crucial not only to comply by the spectral mask regulations and to prohibit interference with adjacent channels as mentioned earlier, but also to avoid saturation of the RX frontend even if it is sufficiently isolated from the fundamental harmonic of the TX signal. As mentioned earlier, we are currently investigating multiple solutions to improve the spurious emission of STM-AM circulators. Our current post-layout simulations indicate that all spurs can be realistically reduced to less than –90 dBc.

## V. CONCLUSION

We presented the first IC implementation of magnetless STM-AM circulators. The proposed circuit relies on switched capacitors instead of varactors to achieve the desired modulation, which reduces the overall size and complexity significantly. We also analyzed it using an in-house numerical method, gaining an insight into how it works and showing its robustness against parasitics and clock non-idealities. We validated the analysis with simulations and measurements using a standard 180 nm CMOS process. The maximum measured IX was larger than 60 dB, achieved without any on-chip calibration or external tuning thanks to the inherent cyclic-symmetry of STM-AM circulators. Nevertheless, poor design of the EDS protection circuitry resulted in significant return losses. This problem was fixed by using three external reconfigurable impedance tuners, thus yielding a RL of 13 dB and an IL of 4.8 dB. Further improvements can be attained by using a more advanced low-loss CMOS technology and by optimizing the chip and the PCB layouts, as will be demonstrated in future publications. Compared to previous discrete implementations of STM-AM circulators, the presented CMOS chip reduces the size by an order of magnitude and occupies a total area of only 36 mm$^2$, thus making it an important step on the quest to enable full-duplex radios in the near future.

## APPENDICES

### A. Eigenvalues of STM-AM Circulators

As explained in [46], differential STM-AM circulators, including the modified current-mode architecture presented in this paper, are quasi-linear time-invariant (QLTI) networks, which can be analyzed using *N*-port parameters. For instance, the *Y*-parameters of such circuits can be written in the following form

$$\bar{\bar{Y}}(\omega) = \begin{bmatrix} Y_{11} & Y_{31} & Y_{21} \\ Y_{21} & Y_{11} & Y_{31} \\ Y_{31} & Y_{21} & Y_{11} \end{bmatrix}. \quad (7)$$

Notice that this matrix is cyclic-symmetric and its eigenvalues can be calculated using $\|\bar{\bar{Y}} - \lambda \bar{\bar{U}}\| = 0$, where $\bar{\bar{U}}$ is the unitary matrix, which results in

$$\lambda_c = Y_{11} + Y_{21} + Y_{31} \quad (8)$$
$$\lambda_+ = Y_{11} + e^{-j2\pi/3} Y_{21} + e^{+j2\pi/3} Y_{31} \quad (9)$$
$$\lambda_- = Y_{11} + e^{+j2\pi/3} Y_{21} + e^{-j2\pi/3} Y_{31} . \quad (10)$$

The eigenvectors associated with (8)-(10) also read

$$\bar{V}_c = [1,1,1]^T \quad (11)$$
$$\bar{V}_+ = \left[1, e^{+j2\pi/3}, e^{-j2\pi/3}\right]^T \quad (12)$$
$$\bar{V}_- = \left[1, e^{-j2\pi/3}, e^{+j2\pi/3}\right]^T . \quad (13)$$

The physical interpretation of these eigenvectors is that they are a unique set of excitations at the circulator's three ports, the superposition of which yields the solution to exciting only one port while terminating the other ports with matched loads.

Therefore, the port currents can in general be written in the following form

$$I_n = I_c + I_+ e^{+j(n-1)2\pi/3} + I_- e^{-j(n-1)2\pi/3} , \quad (14)$$

where the quantities $I_c$, $I_+ e^{+j(n-1)2\pi/3}$, and $I_- e^{-j(n-1)2\pi/3}$ are what we call the common, the clockwise, and the counter clockwise modes, respectively. Also, if the common mode is zero, then (14) reduces to (2).

### B. Composite Floquet Scattering Matrix Method

The *S*-parameters sensitivity analysis presented in Sec. III was performed using an in-house frequency-domain numerical method which we call the composite Floquet scattering matrix method. Compared to commercial circuit simulators, this method expedites the study of parasitic effects and the impact of synchronization errors on the relevant metrics of parametric circuits, thus giving an insight into the robustness of these circuits and allowing to optimize the performance. In this appendix, we briefly describe the underlying principles of this method. In general, the incoming and outgoing waves at each port of an LPTV multi-port network can be written as a superposition of infinite number of harmonics as follows:

$$v^\pm(t) = \sum_{n=-\infty}^{\infty} v_n^\pm e^{j(\omega + n\omega_s)t} , \quad (15)$$

where $v_n^\pm$ are the corresponding Fourier series coefficients, and $\omega_s$ is the modulation frequency. In what follows, we take advantage of the fact that such linear time-periodic network can be treated as a linear *time-invariant* system with infinite number of harmonic ports per each physical port. This allows us to introduce a Floquet scattering matrix (FSM) which for a two-port system with the same real reference impedance $Z_0$ takes the following form

$$\begin{pmatrix} \mathbf{v}_1^- \\ \mathbf{v}_2^- \end{pmatrix} = \begin{pmatrix} \mathbf{S}_{11} & \mathbf{S}_{12} \\ \mathbf{S}_{21} & \mathbf{S}_{22} \end{pmatrix} \begin{pmatrix} \mathbf{v}_1^+ \\ \mathbf{v}_2^+ \end{pmatrix}, \quad (16)$$

where $\mathbf{S}_{ij}$ are square matrices relating the incoming and outgoing power waves at all frequencies of the physical ports *i* and *j*, i.e.,

$$S_{ij,nm} = v_{i,m}^- / v_{j,n}^+ , \quad (17)$$

where *n* and *m* denote the corresponding harmonics in the expansion (15). To obtain the FSM of an arbitrary circuit, the FSMs of each separate element are first aggregated into a single matrix relating the incoming and outgoing power waves at the terminals of all elements, hence,

$$\mathbf{v}_i^- = \sum_j \mathbf{S}_{ij} \mathbf{v}_i^+ . \quad (18)$$

Next, the port indexes are split into two subsets corresponding to inner connections of the network and external ports denoted by small and capital indexes $p$, $q$ and $P$, $Q$, respectively,

$$\mathbf{v}_p^- = \sum_q \mathbf{S}_{pq} \mathbf{v}_q^+ + \sum_Q \mathbf{S}_{pQ} \mathbf{v}_Q^+ \tag{19}$$

$$\mathbf{v}_P^- = \sum_q \mathbf{S}_{Pq} \mathbf{v}_q^+ + \sum_Q \mathbf{S}_{PQ} \mathbf{v}_Q^+ . \tag{20}$$

Since the inner ports are, by definition, interconnected, the following additional relation can be found

$$\mathbf{v}_p^- = \sum_q \mathbf{F}_{pq} \mathbf{v}_q^+ . \tag{21}$$

For example, if only the inner ports 2 and 3 are directly tied together, then we have $\mathbf{v}_2^- = \mathbf{v}_3^+$ and $\mathbf{v}_3^- = \mathbf{v}_2^+$, leading to $\mathbf{F}_{23} = \mathbf{F}_{32} = \mathbf{I}$ with $\mathbf{I}$ being the identity matrix, and the rest of $\mathbf{F}_{pq}$ matrix being zero. Combining (19) and (21), we can find the CFSM relating the power waves only at the external ports as follows

$$\mathbf{v}_P^- = \sum_Q \tilde{\mathbf{S}}_{PQ} \mathbf{v}_Q^+ , \tag{22}$$

where

$$\tilde{\mathbf{S}}_{PQ} = \mathbf{S}_{PQ} + \sum_{p'q'} \mathbf{S}_{Pq'} [(\mathbf{F}_{pq}) - (\mathbf{S}_{pq})]^{-1}_{q'p'} \mathbf{S}_{p'Q} . \tag{23}$$

The CFSM can now be treated as a new black-box element that can be connected to other elements in the same fashion.

ACKNOWLEDGMENT

The authors would like to thank Dr. V. Singh of MaxLinear Inc. and Prof. R. Gharpurey of the University of Texas at Austin for their advice on various technical issues.